\documentclass[12pt]{revtex4}
\usepackage{amssymb,epsf}
\usepackage{latexsym}
\begin{document}

\title{On topological charged braneworld black holes}
\author{Ahmad Sheykhi $^{1,2}$\footnote{sheykhi@mail.uk.ac.ir} and Bin Wang $^{3}$\footnote{wangb@fudan.edu.cn}}
\address{$^1$Department of Physics, Shahid Bahonar University, P.O. Box 76175, Kerman, Iran\\
         $^2$Research Institute for Astronomy and Astrophysics of Maragha (RIAAM), Maragha,
         Iran\\
$^3$  Department of Physics, Fudan University, Shanghai 200433,
China}

\begin{abstract}

We study a class of topological black hole solutions in RSII
braneworld scenario in the presence of a localized Maxwell field
on the brane. Such a black hole can carry two types of charge, one
arising from the extra dimension, the tidal charge, and the other
one from a localized gauge field confined to the brane. We find
that the localized charge on the brane modifies the bulk geometry
and in particular the bulk Weyl tensor. The bulk geometry does not
depend on different topologies of the horizons. We present the
temperature and entropy expressions associated with the event
horizon of the braneworld black hole and by using the first law of
black hole thermodynamics we calculate the mass of the black hole.

\end{abstract}

 \maketitle

In the past years there has been a lot of interest in the
braneworld scenario, based on the assumption that all gauge fields
in standard model of particle physics are confined on a $3$-brane,
playing the role of our $4$-dimensional universe, embedded in a
higher dimensional spacetime, while the gravitational field, in
contrast, is usually considered to live in the whole spacetime.
The first picture appeared in braneworld scenarios was the second
Randall-Sundrum model (RSII) in which, our universe observed as a
positive tension $3$-brane embedded in a $5$-dimensional anti
de-Sitter bulk \cite{RS}. In  this model the localization of
gravity happens on the brane due to the negative bulk cosmological
constant and the cross over between $4$-dimensional and
$5$-dimensional gravity is set by the anti de-Sitter radius.
Within the context of the RSII scenario, it is important that the
induced metric on the brane is, in the low energy regime, the
solution predicted by standard general relativity in four
dimensions. Otherwise the usual astrophysical properties of black
holes and stars would not be recovered. Therefore it is natural to
assume the formation of black hole in the braneworld due to
gravitational collapse of matter trapped on the brane. In fact,
the construction and study of black hole solutions on the brane
has been one of the most important and intriguing challenge in
braneworld physics. There are several reasons why this problem is
so challenging. First, the effective gravitational field equation
on the brane is not the usual Einstein one but contains higher
correction terms due to the nonlocal bulk effects on the brane and
therefore is more complicated compared with the usual
gravitational field equations. Second, even one finds the solution
of the effective gravitational field equations on the brane, one
can not regard it as a braneworld black hole solution. One can
just consider this solution as an initial data for the evolution
of the brane into the bulk. The first attack on this problem was
done by Chamblin, Hawking and Reall who investigated the
gravitational collapse of uncharged, non-rotating matter in RSII
braneworld model \cite{Cham1}. They showed that a static uncharged
black hole on the brane is described by a ``black cigar" solution
in five dimensions. If this cigar extends all the way down to the
anti-de Sitter horizon, then we recover the metric for a black
string in anti-de Sitter spacetime. However, such a black string
is unstable near the anti-de Sitter horizon \cite{Gre1,Gre2}. An
exact braneworld black hole solution satisfies a closed system of
effective gravitational field equations on the brane, describing
an uncharged black hole in the RSII scenario was obtained in
\cite{Dad}. By using the braneworld gravitational field equations
derived in \cite{Shi}, it was shown that a Reissner-Nordstrom
geometry could arise on the brane provided that the bulk Weyl
tensor takes a particular form. The solution in \cite{Dad} carries
a ``tidal charge", arising from the projection of the bulk free
gravitational field effects onto the brane. However, it was argued
in \cite{Cham2} that although the solution in \cite{Dad} was
claimed to describe an uncharged black hole, one can not regard it
as a braneworld black hole solution. One can just consider this
solution as an initial data for the evolution of the brane into
the bulk. Until this evolution is performed and boundary
conditions in the bulk are imposed, it is not clear what this
solution represents. For example, it might give rise to some
pathology such as a naked curvature singularity. Therefore the
main problem remains in the braneworld black hole physics is to
study the effect of the braneworld black hole on the bulk
geometry, and in particular the nature of the off-brane horizon
structure. Indeed, the analytical solution for the bulk spacetime
has not been found until now. The numerical calculations on the
bulk geometry in the case of charged and uncharged braneworld
black holes have been investigated in \cite{Cham2} and
\cite{Shibata}, respectively. Other attempts on the study of
braneworld black holes and their physical properties have been
carried  out in
\cite{Cas1,Cas2,Bro,Gre3,Ali,kof,Bwang,Bwang2,Yosh,Yosh2}.

The purpose of the present Letter is to tackle the first problem
mentioned above in the braneworld black hole physics. We will
consider the Maxwell gauge fields confined onto the brane.
Employing a simple strategy, we solve gravitational field
equations on the brane and obtain the charged topological
braneworld black hole solutions. Our solution is the
generalization of \cite{Cham2} to different horizon topologies. We
also present the temperature and entropy expressions associated
with the event horizon of the braneworld black hole and calculate
the mass of the black hole by using the first law of black hole
thermodynamics. Since the flux lines of gauge fields can pierce
the horizon only when they intersect the brane, our bulk theory is
the same as that of the uncharged case and one might expect that
the ``black cigar" solution still describes the bulk containing
the charged braneworld black hole. Here we will not repeat the
discussion on the bulk metric, since we see that the bulk geometry
does not depend on different topologies of the horizons, thus our
bulk metric is the same as that discussed in \cite{Cham2} for the
spherically symmetric braneworld black hole.

We start with the effective field equations on a 3-brane embedded
in the 5-dimensional anti de-Sitter spacetime with $\mathbb{Z}_2$
symmetry expressed as \cite{Shi}
\begin{eqnarray}\label{eqn}
G_{\mu\nu}=-\Lambda g_{\mu\nu} +8\pi G
T_{\mu\nu}+\kappa_5^4\pi_{\mu\nu}-E_{\mu\nu},
\end{eqnarray}
where
\begin{eqnarray}
G&=&\frac{\kappa^4_5}{48\pi} \lambda,  \hspace{0.7cm}
\Lambda=\frac{\kappa_5^2}{2}\Bigl(
\Lambda_5+\frac{\kappa_5^2}{6}\lambda^2
\Bigr).\\
\end{eqnarray}
Here $\kappa_5$ and $\Lambda_5$ are, respectively, the
five-dimensional gravity coupling constant and cosmological
constant. The factor $\Lambda$ is the effective cosmological
constant on the brane, $\lambda$ is the brane tension, and
$T_{\mu\nu}$ is the stress energy tensor confined onto the brane,
so $T_{AB}\,n^B=0$, where $n^A$ is the unit normal to the brane.
The first correction term relative to Einstein's gravity is the
inclusion of a quadratic term $\pi_{\mu\nu}$ in the stress-energy
tensor, arising from the extrinsic curvature term in the projected
Einstein tensor, and is given by
\begin{equation}
\pi_{\mu\nu}=\frac{1}{12}T T_{\mu\nu}-\frac{1}{4}
T_{\mu\alpha}T_{\nu}^{\ \alpha}{}+
\frac{1}{8}\,g_{\mu\nu}\left(T_{\alpha\beta}T^{\alpha\beta}-\frac{1}{3}T^2
\right) \,.
    \label{inducedEFE}
\end{equation}
The second correction term, ${E}_{\mu\nu}$, is the projection of
the five-dimensional bulk Weyl tensor onto the brane, which is
defined as $ E_{\mu\nu} = {}^{(5)}C_{\mu\alpha\nu\beta} n^\alpha
n^\beta $ and encompasses the nonlocal bulk effect. The only
general known property of this nonlocal term is that it is
traceless, namely ${ E}^{\mu}{}_{\mu}=0$. Using the traceless
property of the projected Weyl tensor, Eq. (\ref{eqn}) can be
simplified into
\begin{equation}\label{R}
R=4\Lambda-8\pi G \,T-
\frac{\kappa_5^4}{4}\left(T_{\alpha\beta}T^{\alpha\beta}-\frac{1}{3}
T^2 \right) \,.
\end{equation}
We would like to find the topological black hole solutions of the
field equations (\ref{eqn}). We assume the induced metric on the
brane in the form
\begin{equation}\label{metric}
ds^2=-f(r)dt^2 +{dr^2\over f(r)}+ r^2d\Omega_{k}^2 ,
\end{equation}
where  $d\Omega_{k}^2$ is the line element of a two-dimensional
hypersurface $\Sigma$ with constant curvature,
\begin{equation}\label{met}
d\Omega_k^2=\left\{
  \begin{array}{ll}
    $$d\theta^2+\sin^2\theta d\phi^2$$,\quad \quad\!\!{\rm for}\quad $$k=1$$, &  \\
    $$d\theta^2+\theta^2 d\phi^2$$,\quad\quad\quad {\rm for}\quad $$k=0$$,&  \\
    $$d\theta^2+\sinh^2\theta d\phi^2$$, \quad {\rm for}\quad $$k=-1$$.&
  \end{array}
\right.
\end{equation}
For $k = 1$, the topology of the event horizon is the two-sphere
$S^2$, and the spacetime has the topology $R^2 \times S^2$. For $k
= 0$, the topology of the event horizon is a torus and the
spacetime has the topology $R^2 \times T^2$. For $k = -1$, the
surface $\Sigma$ is a two-dimensional hypersurface $H^2$ with
constant negative curvature. In this case the topology of
spacetime is $R^2 \times H^2$. It is not necessary to take the
exact metric describing a topological braneworld black hole in the
form (6). In general one may expect that $g_{rr}\neq -
{g_{tt}}^{-1}$. But, it is well known that the induced metric
describing a charged black hole should be close to
Reissner-Nordstrom metric, so our ansatz for the braneworld black
hole metric is a good guess \cite{Cham2}.

Assuming the localized gauge field on the brane is the Maxwell
field with action
\begin{equation}
S = -\frac{1}{16 \pi G} \int d^4 x \sqrt{-g} F_{\mu\nu}
F^{\mu\nu}.
\end{equation}
The corresponding  localized energy-momentum tensor on the brane
can be written as
\begin{equation}\label{Tem}
T_{\mu\nu} = \frac{1}{4\pi G} \left(F_{\mu \rho} F_{\nu}\,^{\rho}
- \frac{1}{4} g_{\mu\nu} F_{\rho\sigma} F^{\rho\sigma} \right).
\end{equation}
which is traceless, satisfying $T=T_{\mu}^{\ \mu}=0$. We also
assume that there is a localized static point charge on the brane
which produces an electric field
\begin{equation}\label{Ftr}
F_{tr}=\frac{q}{r^2},
\end{equation}
where $q$ is the charge parameter. Using metric (\ref{metric}),
the electric field (\ref{Ftr}) and Eq. (\ref{Tem}) for the total
energy-momentum tensor localized on the brane, one can show that
Eq. (\ref{R}) has a solution of the form
\begin{equation}\label{f}
f(r)=k-{\frac {2m}{r}}-\frac{\Lambda}{3}{r}^2+{\frac
{\beta+q^2}{{r}^{2}}}+{\frac {1}{240}}\,{\frac
{{\kappa_5}^{4}{q}^{4}}{{r}^{6}}},
\end{equation}
where  $m$ and $\beta$ are arbitrary integration constants and we
have assumed $4\pi G =1$, for simplicity. Although in \cite{Dad},
$\beta>0$ has been interpreted as a tidal charge associated with
the bulk Weyl tensor, in the presence of localized charge on the
brane, it is quite possible to take $\beta<0$ as pointed out in
\cite{Cham2}. Indeed, the projected Weyl tensor, transmits the
tidal charge stresses from the bulk to the brane. One may also
interpret $\beta$ as a five-dimensional mass parameter
\cite{Cham2}. The horizons can be found by solving Eq. $f(r)=0$.
This equation cannot be solved analytically except for $q=0$. The
event horizon of the charged braneworld black hole locates at
$r_{+}$ where $r=r_{+}$ is the largest root of equation $f(r)=0$.
Inserting solution (\ref{f}) into field equations (\ref{eqn}), we
obtain the components of the five-dimensional bulk Weyl tensor.
The result is
\begin{equation}\label{weyl}
E^{t}_{\ t}=E^{r}_{\ r}=-E^{i}_{\
i}=\frac{\beta}{r^4}+\frac{1}{24}\frac{\kappa_5^4 q^4}{r^8},
\end{equation}
where $i=1,2$. Clearly the traceless nature of the Weyl tensor is
obeyed. Eqs. (\ref{eqn}) with solutions (\ref{f}) and (\ref{weyl})
form a closed system of equations on the brane.

Some discussions on our solution are needed.  In the special case
$k=1$ and $\Lambda=0$, $q=0$, our solution (11) reduces to the
uncharged braneworld black hole solution found in \cite{Dad}. In
the case $k=1$ and $\Lambda=0$, our solution (11) reduces to the
charged black hole solution presented in \cite{Cham2}. With the
presence of the charge on the brane, the bulk geometry has to
change, since now $T_{\mu\nu}\neq0$. In other words, the localized
charge on the brane will induce changes in the bulk geometry and
therefore modifies the bulk Weyl tensor. This property keeps for
different topologies of the horizon. Further from Eq. (\ref{weyl})
we see that the horizon topology of the braneworld black hole does
not affect the bulk geometry and therefore the bulk Weyl tensor is
independent of the constant curvature $k$.

In the following we are going to calculate the conserved and
thermodynamic quantities of the braneworld black hole. We will
adopt a simple strategy based on the profound connection between
gravity and thermodynamics which has recently been revealed in
various gravity theories \cite{Jac}-\cite{Sheywang}, showing the
deep correspondence between the gravitational equation describing
the gravity in the bulk and the first law of thermodynamics on the
apparent horizon. This connection sheds the light on holography
since the gravitation equations persist the information in the
bulk while the first law of thermodynamics on the apparent horizon
contains the information on the boundary. Besides, this connection
was shown as a useful tool to extract the entropy of the
braneworld. In the general case, gravity on the brane does not
obey the Einstein theory and the usual area formula for the black
hole entropy does not hold on the brane. The relation between the
braneworld black hole horizon entropy and its geometry is not
known. It was argued in \cite{Shey1,Shey2} that the entropy
associated with the apparent horizon on the brane can be extracted
from the obtained gravity and thermodynamics correspondence. The
entropy and temperature associated with the apparent horizon of
the FRW universe on the brane, in the RSII braneworld model, are
found in \cite{Cai4,Shey1} with form
\begin{eqnarray}\label{S}
 S&=&\frac{2 \pi \ell
}{G_{5}}{\displaystyle\int^{\tilde r_A}_0\frac{\tilde{r}_A^{2}
}{\sqrt{\tilde{r}_A^2+\ell^2}}d\tilde{r}_A}
=\frac{2\pi{\tilde{r}_A}^{3}}{3G_{5}}
 \times
{}_2F_1\left(\frac{3}{2},\frac{1}{2},\frac{5}{2},
-\frac{{\tilde{r}_A}^2}{\ell^2}\right),\\
T&=&\frac{1}{2\pi {\tilde{r}_A}},\label{T}
\end{eqnarray}
where $\tilde{r}_A$ is the apparent horizon radius and $\ell$ is
the AdS radius of the bulk spacetime which is related to the bulk
cosmological constant. Here ${}_2F_1(a,b,c,z)$ is a hypergeometric
function and $G_{5}=\kappa_{5}^2/8\pi$ is the gravitational
constant in five dimensions. Recently, we have shown that the
extracted apparent horizon entropy, given in Eq. (\ref{S}),
satisfies the generalized second law of thermodynamics
\cite{Sheywang}. The satisfaction of the generalized second law of
thermodynamics further supports that the entropy (13) is a
reasonable thermodynamical entropy describing the brane.

Now we suppose that the temperature and entropy formula (\ref{S})
and (\ref{T}) also hold on the event horizon of the black hole on
the brane. Replacing the apparent horizon radius $\tilde r_A$ by
the black hole horizon radius $r_{+}$, we have the temperature and
entropy on the event horizon of the braneworld black hole
\begin{eqnarray}\label{S2}
S&=&\frac{2 \pi \ell
}{G_{5}}{\displaystyle\int^{r_{+}}_0\frac{r_{+}^{2}
}{\sqrt{r_{+}^2+\ell^2}}dr_{+}}=\frac{2\pi{r_{+}}^{3}}{3G_{5}}
 \times
{}_2F_1\left(\frac{3}{2},\frac{1}{2},\frac{5}{2},
-\frac{r_{+}^2}{\ell^2}\right),\\
T&=&\frac{1}{2\pi r_{+}}.\label{T2}
\end{eqnarray}
Eq. (16) is exactly the Hawking temperature on the event horizon.
The validity of (15) to describe the event horizon entropy of the
braneworld black hole can be justified by considering its limiting
case with $\tilde{r}_{+} \ll\ell$. Physically this limit means
that the size of extra dimension is very large if compared with
the black hole event horizon radius. In this limit Eq. (15)
reduces to the five-dimensional area formula for the black hole
entropy $S =2\Omega_{3}{\tilde{r}_+}^{3}/4G_{5}$, where
$\Omega_{3}=4\pi/3$ is the volume of a unit sphere. The factor $2$
comes from the $\mathbb{Z}_2$ symmetry in the bulk. This is an
expected result since in this regime the anti de-Sitter bulk
reduces to the Minkowski spacetime. And due to the absence of the
negative cosmological constant in the Minkowski bulk, no
localization of gravity happens on the brane. Thus the gravity on
the brane is still five-dimensional and the entropy formula on the
black hole event horizon obeys the five-dimensional area formula
\cite{Shey1}.

Adopting the first law of black hole thermodynamics on the event
horizon $r_{+}$ and considering that the electric charge of black
hole does not affect its mass, we just need to discuss the
uncharged case with the first law
\begin{equation}\label{frst}
dM=TdS.
\end{equation}
Integrating (\ref{frst}) and inserting (\ref{S2}) and (\ref{T2}),
we obtain the mass of the braneworld black hole
\begin{equation}\label{mass2}
M=\frac{\ell}{G_{5}}\int^{r_{+}}_0 \frac{ r_{+}d
r_{+}}{\sqrt{r_{+}^2+\ell^2}}=\frac{\ell}{G_{5}}\left(\sqrt{r_{+}^2+\ell^2}-\ell
\right).
\end{equation}
It is interesting to see that in the limiting case $\tilde{r}_{+}
\ll\ell$, the mass formula (\ref{mass2}) reduces to
\begin{equation}\label{mass3}
M=\frac{r_{+}^2}{2 G_{5}},
\end{equation}
which is exactly the mass of the five-dimensional black hole in
Einstein gravity.

In conclusion, we have obtained a class of topological black hole
solutions in RSII braneworld scenario in the presence of a
localized Maxwell field on the brane. We have shown that the
localized charge on the brane modifies the bulk geometry and in
particular the bulk Weyl tensor. The horizon topology of the
braneworld black holes does not affect the geometry of extra
dimension. We presented the temperature and entropy expressions
associated with the event horizon of the braneworld black hole. We
also obtained the mass of the braneworld black holes through the
use of the first law of black hole thermodynamics.

We would like to mention here that in this Letter we have not
studied fully the effect of the braneworld black hole on the bulk
geometry, and in particular the nature of the off-brane horizon
structure. This has been done for solutions which reduce to the
Schwarzschild black hole on the brane \cite{Cham1}. We have
adopted a different approach: instead of starting from an induced
metric on the brane, we have solved the closed system of the
effective field equations for the induced metric on the brane in
RSII model, and found a  class of topological braneworld black
holes. Therefore the main problem remains to find the exact bulk
metric that describes a topological braneworld black hole. This
was solved for uncharged black holes in three dimensions
\cite{Emp}. Unfortunately, the higher dimensional generalization
of this metric is still not known. In general the bulk spacetime
may be given, by solving the full five-dimensional equations, and
the geometry of the embedded brane is then deduced. Due to the
complexity of the five-dimensional equations, one may follow the
strategy outlined in this Letter, by considering the intrinsic
geometry on the brane, which encompasses the imprint from the
bulk, and consequently evolve the metric off the brane. However,
in this Letter we did not study the effects of the braneworld
black hole on the bulk geometry, and in particular the nature of
the topological horizon structure in the bulk. Indeed, determining
the bulk geometry is an extremely difficult task which needs
numerical calculations, so it was not explored here.

\acknowledgments{This work has been supported financially by
Research Institute for Astronomy and Astrophysics of Maragha,
Iran. The work of B. W. was support in part by NNSF of China,
Shanghai Education Commission and Shanghai Science and Technology
Commission. }


\begin{thebibliography}{99}
\bibitem{RS} L. Randall, R. Sundrum, Phys. Rev. Lett. {\bf 83}, 4690
(1999).
\bibitem{Cham1}
A. Chamblin, S.W. Hawking and H.S. Reall, Phys. Rev. D {\bf 61},
065007 (2000).

 \bibitem{Gre1} R. Gregory and R. Laflamme, Phys. Rev. Lett. {\bf70}, 2837
 (1993).
\bibitem{Gre2} R. Gregory, Class. Quantum Grav. {\bf17}, L125 (2000).

\bibitem{Dad} N. Dadhich, R. Maartens, P. Papadopoulos and V. Rezania,  Phys. Lett. B {\bf487}, 1 (2000).

\bibitem{Shi}
T.~Shiromizu, K.~Maeda and M.~Sasaki,  Phys. Rev. D {\bf62},
024012 (2000).

\bibitem{Cham2}A. Chamblin, H. S. Reall, H. Shinkai and T. Shiromizu, Phys. Rev. D {\bf 63}, 064015 (2001).

\bibitem{Shibata}
T. Shiromizu and M. Shibata, Phys. Rev. D {\bf62}, 127502 (2000).

\bibitem{Cas1} R. Casadio, A. Fabbri and L. Mazzacurati, Phys. Rev. D {\bf 65}, 084040
(2002).
\bibitem{Cas2} R. Casadio and L. Mazzacurati, Mod. Phys. Lett. A {\bf18}, 651
(2003).
\bibitem{Bro} K. A. Bronnikov, H. Dehnen, V. N. Melnikov, Phys. Rev. D {\bf 68},
024025 (2003).
\bibitem{Gre3} R. Gregory, R. Whisker, K. Beckwith and C. Done, JCAP {\bf0410},
013 (2003).
\bibitem{Ali} A. N. Aliev, A. E. Gumrukcuoglu,  Phys. Rev. D {\bf 71}, 104027
(2005).
\bibitem{kof}
G. Kofinas, E. Papantonopoulos and V. Zamarias, Phys. Rev. D {\bf
66}, 104028 (2002).

\bibitem{Bwang}  L.H. Liu, B. Wang, G. H. Yang, Phys. Rev. D {\bf 76}, 064014
(2007).
\bibitem{Bwang2} J. Shen, B. Wang, R. K. Su, Phys. Rev. D {\bf 74}, 044036
(2006).

\bibitem{Yosh}  H. Kudoh, T. Tanaka and T. Nakamura, Phys. Rev. D {\bf68}, 024035
(2003).
 \bibitem{Yosh2} H. Yoshino, JHEP {\bf0901}, 068 (2009).


\bibitem{Jac} T. Jacobson, Phys. Rev. Lett. {\bf75}, 1260 (1995).

\bibitem{Elin} C. Eling, R. Guedens, and T. Jacobson,
Phys. Rev. Lett. {\bf96}, 121301 (2006).
\bibitem{Cai1} M. Akbar and R. G. Cai, Phys. Lett. B {\bf635}, 7 (2006)
.
\bibitem{CaiAk} M.~Akbar and R.~G.~Cai,
   Phys. Lett. B {\bf648}, 243 (2007).

\bibitem{Pad1}T. Padmanabhan, Class. Quant. Grav. {\bf 19}, 5387
(2002).
 \bibitem{Pad2} T. Padmanabhan, Phys. Rept. {\bf 406}, 49 (2005).
 \bibitem{Pad3} A.~Paranjape, S.~Sarkar and T.~Padmanabhan,
    Phys.\ Rev.\ D {\bf 74}, 104015 (2006).
   \bibitem{Pad4} T.~Padmanabhan and A.~Paranjape,
  Phys. Rev. D {\bf75} (2007) 064004.
  \bibitem{Wu}  S. F. Wu, B. Wang, and G. H Yang, Nucl. Phys. B {\bf799} (2008) 330.

  \bibitem{Cai2} M.~Akbar and R.~G.~Cai, Phys. Rev. D {\bf 75}, 084003 (2007).
  \bibitem{Cai3} R.~G.~Cai and L.~M.~Cao, Phys.Rev. D {\bf 75}, 064008
  (2007).

\bibitem{CaiKim} R. G. Cai and S. P. Kim, JHEP {\bf0502}, 050
(2005).
\bibitem{cao}  R. G. Cai, L.M. Cao, Y.P. Hu, arXiv: 0809.1554.

 \bibitem{Fro} A. V. Frolov and L. Kofman, JCAP {\bf 0305},
009 (2003).
\bibitem{Dan} U. K. Danielsson, Phys. Rev. D {\bf71},
023516(2005) .
\bibitem{Boss} R. Bousso, Phys. Rev. D {\bf71}, 064024 (2005).
\bibitem{Cal} G. Calcagni, JHEP {\bf0509}, 060 (2005).
\bibitem{Wang} B. Wang, E.
Abdalla and R. K. Su, Phys.Lett. B {\bf503},  394 (2001).
\bibitem{wang} B. Wang,
E. Abdalla and R. K. Su, Mod. Phys. Lett. A {\bf17},  23 (2002).
\bibitem{RGcai} R.~G.~Cai and Y.~S.~Myung, Phys.\ Rev.\ D {\bf 67}, 124021 (2003).
\bibitem{Cai4} R.~G.~Cai and L.~M.~Cao,
  Nucl. Phys. B {\bf785} (2007) 135.

\bibitem{Shey1} A. Sheykhi, B. Wang and R. G. Cai, Nucl. Phys. B {\bf
779} (2007)1.
  \bibitem{Shey2} A. Sheykhi, B. Wang and R. G. Cai, Phys. Rev. D {\bf
76} (2007) 023515.

\bibitem{Sheywang0} A. Sheykhi, B. Wang, arXiv:0811.4477.
\bibitem{Sheywang} A. Sheykhi, B. Wang, arXiv:0811.4478.

\bibitem{Emp}
R.~Emparan, G.~T.~Horowitz, and R.~C.~Myers, JHEP {\bf 0001}, 007
(2000).
\end{thebibliography}
\end{document}